%% file: GW_Overview_EP_GA.tex
\documentclass[master]{ws-rv9x6} 

\usepackage{subfigure}   
\usepackage{ws-rv-thm}   
\usepackage{ws-index}     
\usepackage{booktabs}	 
\usepackage{bm}
\usepackage{color}
\usepackage[square]{ws-rv-van}   
\usepackage[colorlinks]{hyperref}
\usepackage[final ]{pdfpages}  
\newindex{aindx}{adx}{and}{Author Index}       

\usepackage{import} 

\makeindex 

\definecolor{darkblue}{rgb}{0,0,0.3}
\hypersetup{citecolor=darkblue}
\hypersetup{linkcolor=darkblue}

\newcommand{\be}{\begin{equation}}
\newcommand{\ee}{\end{equation}}
\newcommand{\beq}{\begin{equation}}
\newcommand{\eeq}{\end{equation}}

\newcommand{\eff}{_{\text{eff}}}
\newcommand\stxt[1]{_{\text{#1}}} 

\newcommand{\Hz}{\ \text{Hz}}
\newcommand{\kHz}{\ \text{kHz}}

\newcommand{\sqrtHz}{/\sqrt{\text{Hz}}}

\begin{document}



\setcounter{page}{1}









\include{Geiger/geiger}

\printindex[aindx]           
\printindex                  

\end{document}

%% file: Geiger/geiger.tex






\chapter{Future Gravitational Wave Detectors Based on Atom Interferometry}\label{sec:ch8}
\author[Remi Geiger]{Remi Geiger}
\address{LNE-SYRTE, Observatoire de Paris, PSL Research University, \\CNRS, Sorbonne Universit\'es, UPMC Univ. \\Paris 06, 61 avenue de lÕObservatoire, 75014 Paris, France.}

\bigtoc

\tableofcontents

\body

\newpage

\begin{abstract}
\textit{\textbf{Abstract}: We present the perspective of using  atom interferometry for  gravitational wave (GW) detection in the mHz to about 10 Hz frequency band. We focus on light-pulse atom interferometers which have been subject to intense developments in the last 25 years. 
We calculate the effect of the GW on the atom interferometer  and present in details the atomic gradiometer configuration which has retained more attention recently. The principle of such a detector is to use free falling atoms to measure the phase of a laser, which is modified by the GW.
We highlight the potential benefits of using atom interferometry compared to optical interferometry as well as the challenges which remain for the realization of an atom interferometry based GW detector. 
We present some of the important noise sources which are expected in such detectors and strategies to cirucumvent them. Experimental techniques related to cold atom interferometers are briefly explained. We finally present the current progress and projects in this rapidly evolving field.}
\end{abstract}

\section{Introduction}
\label{sec:introduction}

Matter-wave interferometry relies on the wave nature of massive particles to realize an interferometer, in analogy with optical interferometry which exploits the wave nature of photons \cite{Berman1997}. 
Several illustrations of matter-wave interference phenomena  have been demonstrated in the 20th century including interference with electron Cooper pairs \cite{Zimmerman1965}, neutrons \cite{Rauch1974}, atoms \cite{Keith1991} or molecules \cite{Borde1994}.
Atom interferometry has benefited from the important progress in the field of cold atom physics which started in the 1980's. The first proofs of principle of atom inteferometers (AIs) in 1991 \cite{Keith1991,Riehle1991,KasevichChu1991,Carnal1991,Robert1991} triggered significant experimental developments.
In particular, the possibility to extend the Ramsey sequence with optical fields to build an atom interferometer based on the atomic recoil
triggered many experiments \cite{Borde1989}.
 The experimental developments led to the realization of different types of AIs adressing several applications, such as inertial sensing \cite{Canuel2006,Geiger2011}, precision measurements of fundamental constants \cite{Bouchendira2011,Rosi2014},  fundamental physics \cite{Lepoutre2012,Zhou2015}, and gravimetry \cite{Gillot2014,Freier2015}.

Parallel to the emergence of concepts for gravitational wave (GW) detectors based on laser interferometry,  studies were conducted  on the effect of gravitational fields in matter-wave interferometry in the 1970s \cite{Linet1976,Stodolsky1979}, and later in  laser spectroscopy \cite{Borde1983}. 
Less than 10 years after the pioneering experiments of 1991, the field of atom interferometry had already been subject to an important progress, for example with the demonstration of a sensitive atomic gravimeter \cite{Peters2001} and gyroscope \cite{Gustavson2000}, and owing to the rapid development of cold atom physics. Such progress motivated reconsidering the application to GW detection.

To this end, various theoretical frameworks were proposed, such as a generalization of the Klein-Gordon equation \cite{Cai1989}, a generalization of the Dirac equation in curved space-time \cite{Borde1994}, or a generalization of the ABCD matrix formalism of optics to matter-wave propagation \cite{Borde2001}.
In 2004, Chiao and Speliotopoulos publish a paper  where they analyze the sensitivity of a matter-wave interferometer using atomic beams emanating from a supersonic atomic source, and claim favorable sensitivities for such devices compared to space-based GW detectors based on laser interferometry \cite{Chiao2004}. Their paper is however subject to debate \cite{Foffa2006,Roura2006}, the various authors finding different results because they  studied different physical experiments. As analyzed in \cite{Delva2007}, the important aspect of such studies lies in the interpretation of the coordinate systems and of the boundary conditions in order to obtain the same result for various descriptions of the experiment, which is at the basis of general relativity.

Rapidly, several teams start to work on the estimation of the sensitivity of AIs for GW detection. 
Delva \textit{et al} \cite{Delva2006}, and Tino and Vetrano (following the work of Bord\'e) \cite{Tino2007}  find the same result for the GW induced phase shift in an AI :
\beq
\Delta\phi\sim h\stxt{GW} L/\lambda\stxt{dB},
\label{eq:phase_shift1}
\eeq
where $h\stxt{GW}$ is the strain amplitude of the GW and $L$ is the physical separation between the two arms of the interferometer.  
The signal is inversely proportional to the de Broglie  wavelength of the atom of mass $m$, $\lambda\stxt{dB}\sim h/mv_L$, with $v_L$ being the velocity of the atoms enterring the interferometer and $h$ the Planck constant. 
If we consider the case of a light-pulse AI, where the splitting of the atomic wave is performed on a light standing wave (\ref{sec:light_pulse_AI} will present this AI in details), then the physical separation between the two interferometer arms is given by $L=T\hbar k/m $, with $k=2\pi/\lambda\stxt{laser}$ the optical wavevector of the light grating, $T$ the time spent by the atoms in the interferometer region and $\hbar=h/2\pi$. Eq.~\eqref{eq:phase_shift1} then simplifies in 
\beq
\Delta\phi\sim h\stxt{GW} \frac{T v_L}{\lambda\stxt{laser}},
\label{eq:phase_shift2}
\eeq
illustrating that it is favorable to use beams of fast atoms and rather long devices (to increase the interrogation time $T$). 

 The authors discuss the comparison with optical interferometers where the phase shift can be writen in a similar form as Eq.~\eqref{eq:phase_shift1} \cite{Delva2006}. The interpretation is then the following: the AI can potentially be more sensitive than the optical interferometer for the same linear dimension and level of phase noise, because the wavelength of the wave is much shorter for atomic waves ($\lambda\stxt{dB}\sim$~pm for beams at few $100 \ \text{m.s}^{-1}$) than for optical waves ($\lambda\stxt{light}\sim$~hundreds of nm). However, the AI cannot compete with optical interferometers for two reasons: \textit{(i)} it is very difficult to realize an interferometer with arm separations $L$ exceeding the meter scale (compared to kilometer scale optical ones); \textit{(ii)} the  flux $\dot{n}$ which determines the detection noise of the interferometer given by $\sigma_\phi\sim1/\sqrt{\dot{n}}$ is more than 10 orders of magnitude smaller for atomic sources than for photons in a laser. Besides these considerations, these papers do not analyze the other noise sources which could limit the sensitivity of the detector.

The strategy for detecting GWs with AIs evolves with a paper of 2008 \cite{Dimopoulos2008PRD}. In this article, the authors formalize the idea of using the AI to read the phase of a laser which is modified by the GW, in analogy with the mirrors which are used as phase references in optical GW detectors. To reject the position noise which degrades the sensitivity of laser interferometers, they propose a gradiometer configuration with two distant AIs interrogated by the same laser beam. 
It is this idea of the gradiometer configuration which has then retained much attention and  which we will describe in details in this chapter.

\section{Detector based on two distant light-pulse atom interferometers}
\label{sec:gradiometer}

\subsection{Principle of the light-pulse atom interferometer}
\label{sec:light_pulse_AI}

\begin{figure}[!h]
\centering
\includegraphics[width=\linewidth]{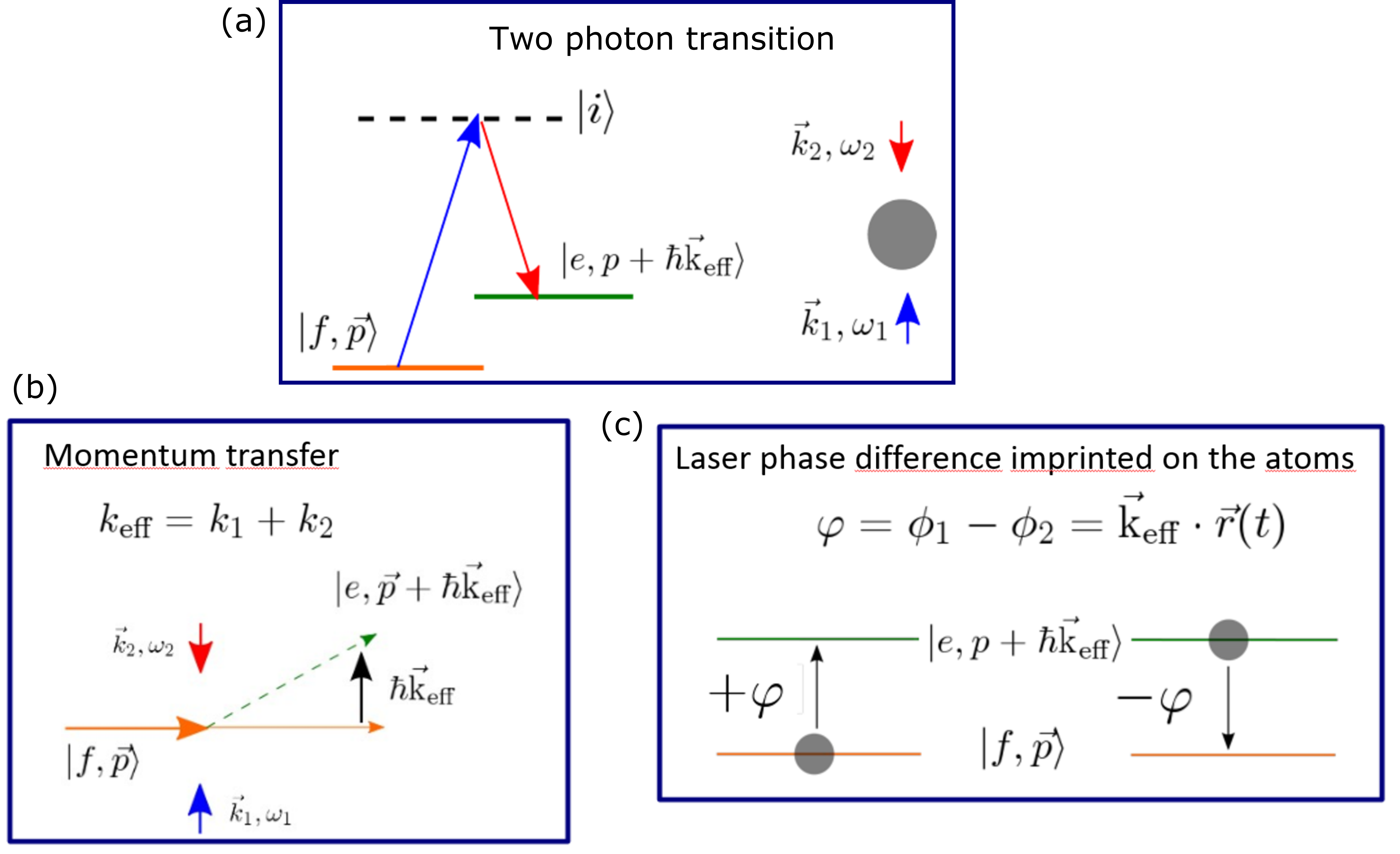}
\caption{(a) Stimulated two photon transition  coupling two momentum states in a 3-level atom. (b) The two-photon transition enables to create quantum superpositions of the two momentum states and thus to create two arms in an interferometer. (c) During the diffraction process, the relative phase $\varphi=\varphi_1-\varphi_2$ between the two lasers is imprinted on the diffracted part of the atomic wavefunction.}
\label{fig:diffraction}
\end{figure}

We start the description of a GW detector based on atom interferometry by describing the origin of the phase shift in a light-pulse AI, where the atomic waves are diffracted on a light grating.
The diffraction process is represented in Fig.~\ref{fig:diffraction}.
 Various schemes exist to realize this diffraction, such as two photon Raman transitions \cite{KasevichChu1991} or Bragg diffraction \cite{Giltner1995}. In both schemes, the atom absorbs a photon form one beam (momentum $\hbar k_1$) and stimulatedly emit a photon from the other beam (momentum $\hbar k_2$). If the two beams are counterpropagating, the momentum of the atom changes by $\hbar(k_1 +k_2)\equiv \hbar k\eff$ which is about twice the optical wavector of the electromagnetic fields used in the process ($k\eff\simeq 2 k_1$). 
Such two photon transition enables to create quantum superpositions of  two momentum states $|\vec{p}\rangle$ and $|\vec{p}+\hbar\vec{k}\eff\rangle$ of the atom, which represent the 2 arms of the interferometer. For optical transitions in atoms, the typical momentum transfer is of the order of 1 cm/s.
Besides the physical separation, the relative phase $\varphi=\varphi_1-\varphi_2$ between the two lasers is imprinted on the diffracted part of the atomic wavefunction \cite{Storey1994}.

\begin{figure}[!h]
\centering
\includegraphics[width=0.8\linewidth]{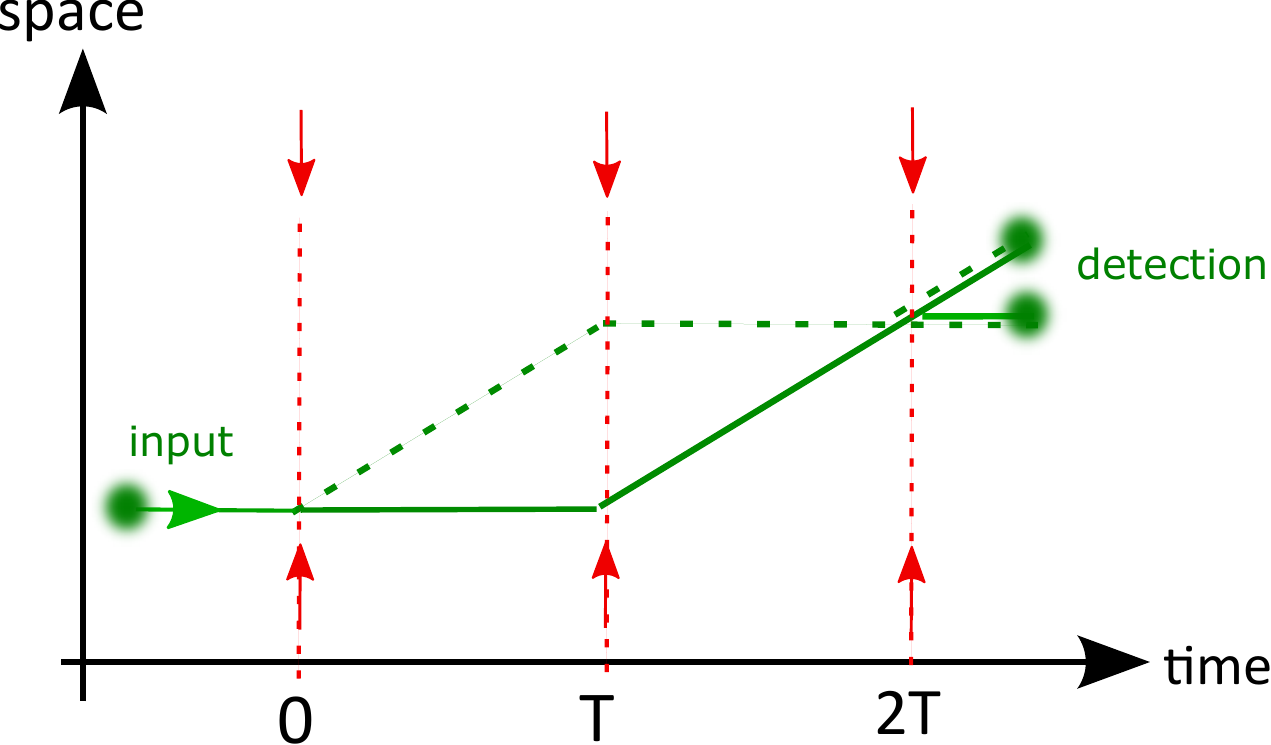}
\caption{ Space-time diagram of the light-pulse atom interferometer (AI). The red arrows represent the two counterpropagating lasers  and the green lines represent the path of the atomic waves.}
\label{fig:AI}
\end{figure}

In analogy with the optical Mach-Zehnder interferometer, it is possible to realize an AI consisting of three light pulses which respectively split, redirect and recombine the atomic wave (see Fig.~\ref{fig:AI}). The three light pulses thus act similarly as the beam splitters and mirrors in an optical interferometer. The beam splitter or mirro condition can be obtained by varying the interaction strength of the laser-atom interaction (e.g. by varying the light pulse duration). 
The phase difference between the two arms can be computed from the relative laser phases imprinted at the different times on the atom  and reads $\Delta\phi=\varphi(0)-2\varphi(T)+\varphi(2T)$, with $T$ the time between the light pulses \cite{Storey1994}. As an illustration, in the case where the AI is used as a gravimeter, the light beams propagate parallel to the local gravity acceleration and the AI phase is given by $\Delta\phi=\vec{k}\eff \cdot \vec{g}T^2$. This allows  to perform absolute measurements of the gravity acceleration at the $10^{-9}$ relative precision level \cite{Peters2001,Gillot2014,Freier2015}.

\subsection{Atom interferometer phase shift in the presence of a GW}
\label{sec:phase_shift}

We present here a derivation of the light-pulse AI phase shift in the presence of a GW. We will concentrate on the effect of the GW on the phase of the lasers, which is impinted on the atomic wavepacket at the diffraction events.
We will not consider the effect of the GW on the phase of the atomic waves themselves (see Eq.~\eqref{eq:phase_shift1}), as it is negligible compared to the laser induced phase in the  configuration which will be considered. We refer, e.g., to Ref.~\cite{TinoVetrano2011} [Eq. (24)] for the full expression with both  the laser induced phase and the AI phase contributions. We will come back to this approximation in the conclusion \ref{sec:conclusion}.

\begin{figure}[!h]
\centering
\includegraphics[width=\linewidth]{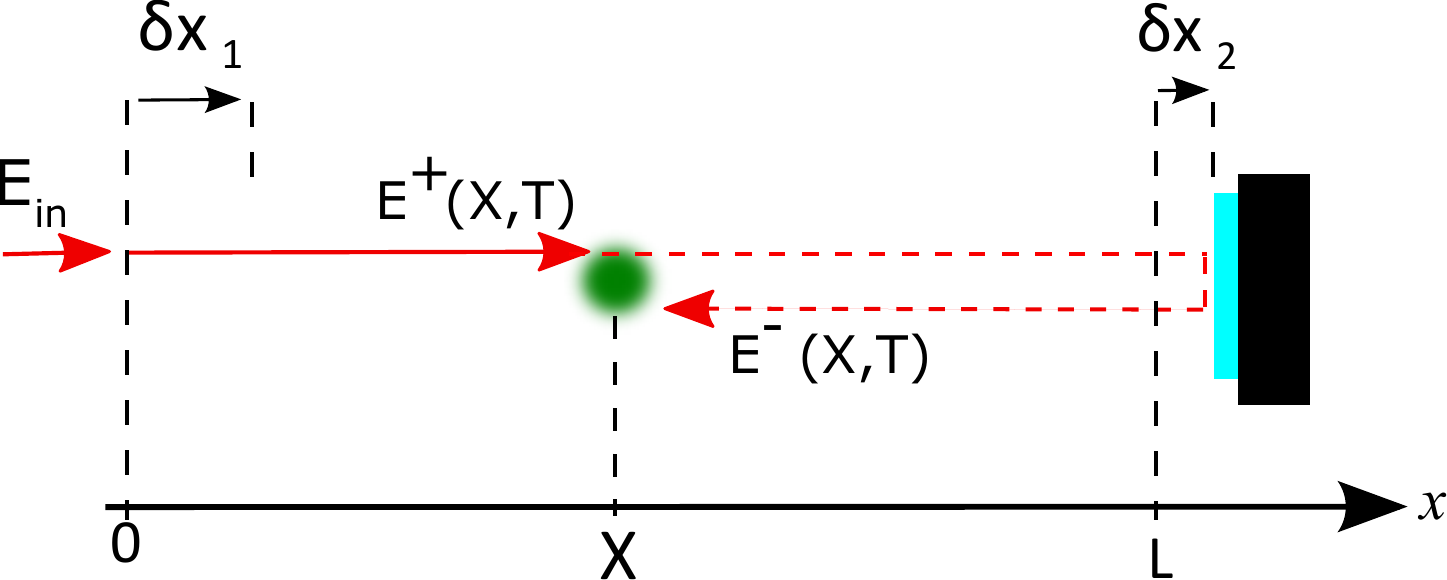}
\caption{ Laser interrogation scheme and notations used in the calculation of the AI phase shifts. $\delta x_1(t)$ and $\delta x_2(t)$ are  the position fluctuations of the input optics and of the retroreflecting mirror with respect to the fixed baseline $L$, respectively.}
\label{fig:scheme_lasers}
\end{figure}

The scheme of the laser interrogation  is shown in Fig.~\ref{fig:scheme_lasers}, where the laser beam is retroreflected.
We remind that the phase difference between the two arms in the AI essentially originates from the local phase of the lasers which is imprinted onto the diffracted wave-packet at the interaction points \cite{Storey1994,Borde2004}. Therefore, the calculation of the AI phase reduces to the calculation of the laser phase of the two counterpropagating beams. We will use the Einstein Coordinates to describe the experiment, where the  GW affects the propagation of light and the atoms are freely falling, i.e. used as phase discriminators. The same result is obtained when considering a different coordinate system \cite{Borde2004,TinoVetrano2011}.

\paragraph{(a) calculation of the phase of the lasers.}
We decompose the electromagnetic field as a superposition of two counterpropagating waves, $E^{\pm}(t)$, respectively propagating towards positive and negative $x$ (see Fig.~\ref{fig:scheme_lasers}). The AI phase shift is determined by the \textit{relative} phase between $E^{+}(t)$ and $E^{-}(t)$, which is  imprinted on the atoms at position $X$. We will determine this relative phase as a function of the mirror position, the frequency of the laser, and the gravitational wave (GW) amplitude.

We consider the effect of a GW propagating in the direction perpendicular to the plane of Fig.~\ref{fig:scheme_lasers} and polarized along the laser propagation direction ($x$). We describe the propagation of electromagnetic (EM) waves in the $x$ direction.
The relativistic invariant is given by \cite{VIRGOphBook}:
\beq
ds^2 = c^2dt^2-dx^2 + h(t)dx^2 =0, 
\eeq
where $h(t)$ is the amplitude of the GW.
For weak GW, the solution is
\beq
dx = \pm [1+\frac{1}{2} h(t)] cdt,
\eeq
where the plus (minus) sign corresponds to the light propagating from left to right (right to left).
The EM wave  is emitted at time $t^+$ from the left input optics (position $\delta x_1(t^+)$) and  arrives at time $T$ at the position $X$, where it interacts with the atoms. The emission time $t^+$ is given by:
\beq
\int_{\delta x_1(t^+)}^X dx = c[T-t^+] +\frac{c}{2} \int_{t^+}^T h(t)dt. 
\eeq
Using perturbation theory with $t^+\approx T-X/c$, we obtain at first order in $\delta x_1,h$:
\beq
t^+ \approx T -\frac{X-\delta x_1(T-\frac{X}{c})}{c} + \frac{1}{2} \int_{T-\frac{X}{c}}^T h(t)dt.
\label{eq:tplus}
\eeq
In a similar way, we obtain the emission time $t^-$ of the EM wave which propagates to the right mirror, is reflected, and propagates back in opposite direction to arrive at position X at time $T$, where it interacts with the atoms. 
Taking into account the propagation from the right mirror to the atoms, $t^-$ is given  by:
\begin{multline}
t^- = T -\frac{1}{c}[2L - X +2\delta x_2(T-\frac{L-X}{c}) -\delta x_1(T-\frac{2L-X}{c})] \\
+ \frac{1}{2} \int_{T-\frac{2L-X}{c}}^T h(t)dt.
\label{eq:tminus}
\end{multline}
At the space-time event $(X,T)$, the atoms  interact with two counterpropagating fields $E^\pm (X,T)$, which we define by
\beq
E^\pm(X,T) \equiv E(t^\pm). 
\eeq
The relative phase $\Delta\varphi=\varphi^+-\varphi^-$ imprinted on the atoms during the atom diffraction is thus determined by the time delay $t^+ - t^-$ between the two emission events, which is obtained from Eqs.~\eqref{eq:tplus} and \eqref{eq:tminus} .

We now consider only slow fluctuations of $\delta x_i(t)$ and $h(t)$ corresponding to frequencies $\omega/2\pi\ll (2L/c)^{-1}$. In particular, we neglect the position fluctuations on a time-scale smaller than the light round-trip time from the atoms to the retroreflecting mirror.
This condition can be ensured by the use of a dedicated suspension system of the optics (see Chapter \ref{sec:ch4})].
In this approximation the fluctuations in Eqs. \eqref{eq:tplus} and \eqref{eq:tminus} are evaluated at time $T$ and become:
\begin{eqnarray}
t^+ & \approx & T -\frac{X}{c} + \frac{\delta x_1(T)}{c}+ \frac{X}{2c}  h(T) \\ 
\label{eq:tplus2}
\bigskip
t^- & \approx & t^+  -\frac{2(L-X)}{c} - \frac{2\delta x_2(T)}{c}+ \frac{2(L-X)}{2c}  h(T).  
\end{eqnarray}

\bigskip
To account for laser phase noise, we write the  EM field  as
\beq
\label{eq:def_field}
E(t)=E_{in}(t) e^{i\alpha(t)}
\eeq
where $E_{in}(t)$ is the amplitude of the  EM field just after the input optics and
\beq
\alpha(t)=2\pi\nu_0 t +\tilde{\phi}(t)
\label{eq:alpha_of_t}
\eeq
is the laser phase. Here $\nu_0$ is the injection laser frequency and $\tilde{\phi}(t)$ is the laser  phase noise.
Assuming that the phase noise  is small and slowly varying, we express it as
\beq
\tilde{\phi}(t+\Delta t)\approx \phi_0+2\pi\delta\nu(t)\Delta t,
\label{eq:model_phase_noise}
\eeq
where $\delta\nu(t)$ is the  frequency noise of the laser. 
This approximation is valid as long as $\Delta t$ is smaller than the typical inverse bandwidth of the noise, meaning that in a sufficiently small region of time around $t$, the phase  is proportional to the instantaneous frequency ($\nu_0+\delta\nu(t)$) of the laser field.

With this model for the phase noise, we obtain the following expression for the EM field at the  point of interaction with the atoms:
\beq
E^+(X,T)\equiv E(t^+) = E_{in}(t^+) e^{i\alpha(t^+)}
\label{eq:Ec_plus},
\eeq
with 
\beq
\alpha(t^+)\approx 2\pi\nu_0 t^+ +\phi_0 +2\pi\delta\nu(T)(t^+-T).
\eeq
Using the above equation for $t^+$, we thus obtain: 
\beq
E^+(X,T) \approx E_{in}(T) e^{2i\pi T[\nu_0 +\delta\nu(T)] +i\phi_0} e^{i\varphi^+(X,T)}
\eeq 
with
\beq
\varphi^+(X,T)=-\frac{2\pi \nu_0}{c}X+\frac{2\pi\nu_0}{c}\delta x_1(T)  + \frac{2\pi}{c}[-\delta\nu(T)+\frac{\nu_0}{2} h(T) ]X.
\label{eq:phi_P}
\eeq
From now on, we will omit the time argument $(T)$ in the variables $\{\delta x_i,\delta\nu,h\}$ for clarity of the equations.
A similar calculation for the $E^-$ field yields the phase
\beq
\varphi^-(X)=-\frac{2\pi \nu_0}{c}(2L-X) +\frac{2\pi\nu_0}{c}[\delta x_1 - 2\delta x_2]  + \frac{2\pi}{c}[-\delta\nu+\frac{\nu_0}{2} h ](2L-X).
\label{eq:phi_M}
\eeq
The relative phase $\Delta\varphi=\varphi^+-\varphi^-$ imprinted on the atoms during the  diffraction is thus:
\beq
\Delta\varphi(X)=2k\left[(L-X)+\delta x_2  + \left[\frac{\delta\nu}{\nu_0}-\frac{h}{2} \right](L-X)\right],
\label{eq:delta_phi_laser}
\eeq
where $k=2\pi\nu_0/c$ is the laser wavevector.
In the retroreflecting configuration, the position noise $\delta x_1$ of the input optics is common to both beam and  is therefore not present in Eq.~\eqref{eq:delta_phi_laser}.

\paragraph{(b) Sensitivity function of the atom interferometer.}
The  AI phase is  determined by the relative phase of the EM fields given by Eq.~\eqref{eq:delta_phi_laser}, and by the sensitivity function $s(t)$ of the three light-pulse AI. 
The formalism of the sensitivity function  was introduced in the context of atomic clocks to  describe the response of an atom interferometer to fluctuating phase contributions (the Fourier transform of $s(t)$ corresponds to the transfer function of the AI). A description of the formalism as well as a measurement of the sensitivity function can be found in \cite{Cheinet2008IEEE}.

Besides the sensitivty function, AIs operate sequentially and deliver a measurement every cycle of duration $T_c=T\stxt{prep}+2T+T\stxt{det}$ during which the atoms are prepared (e.g. laser cooled during a period $T\stxt{prep}$), interrogated in the AI (duration $2T$) and detected at the AI output (duration $T\stxt{det}$).
The AI output signal at cycle $m$ is then given by the convolution product 
\beq
s_\varphi(X,mT_c)=\Delta\varphi(X,t)\otimes s(t-mT_c),
\label{eq:sensitivity_function}
\eeq
with $s(t)\approx\delta(t-2T)-2\delta(t-T)+\delta(t)$ and $\delta(t)$ the Dirac distribution. 
For simplification,  we neglected in this expression of the sensitivity function the finite duration of the light pulse (the full expression  can be found in \cite{Cheinet2008IEEE}).
This approximation corresponds to neglecting the phase fluctuations (e.g. due to $\{\delta x_i(t),\delta\nu(t),h(t)\}$)  of frequencies higher than the Rabi frequency of the two-photon transition, which typically lies in the tens of kHz range. 

\paragraph{(c) Full expression of the AI phase.}
Using Eqs.~\eqref{eq:sensitivity_function} and \eqref{eq:delta_phi_laser}, the output signal of the AI at cycle $m$ reads 
\beq
\label{eq:phase_AI}
s_\varphi\left(X,mT_c\right) = 2 k \left[-\delta x\left(X,t\right) +\delta x_2(t) + \left(\frac{\delta\nu(t)}{\nu_0}-\frac{h(t)}{2}\right)\left(L-X\right) \right]\otimes s(t-mT_c).
\eeq
Here, $\delta x(X,t)$ represents the motion of the  atoms along the laser beam direction due to the fluctuations of the local gravitational acceleration.
This contribution corresponds to the first term $2k(L-X)$ appearing in Eq.~\eqref{eq:delta_phi_laser} which we rewrote as follows for more clarity: \textit{(i)} as $L$ is a constant, it does not contribute to the AI signal and disapears ; \textit{(ii)} to highlight the fact that $X$ might fluctuate because of temporal variations of the local gravitational field in the $x$ direction, we change for the notation $\delta x(X,t)$. We will focus on this contribution in the section on gravity gradient noise reduction \ref{sec:rejection_NN}.

\subsection{Gradiometer configuration.}
\label{par:gradio}

\begin{figure}[!h]
\centering
\includegraphics[width=0.8\linewidth]{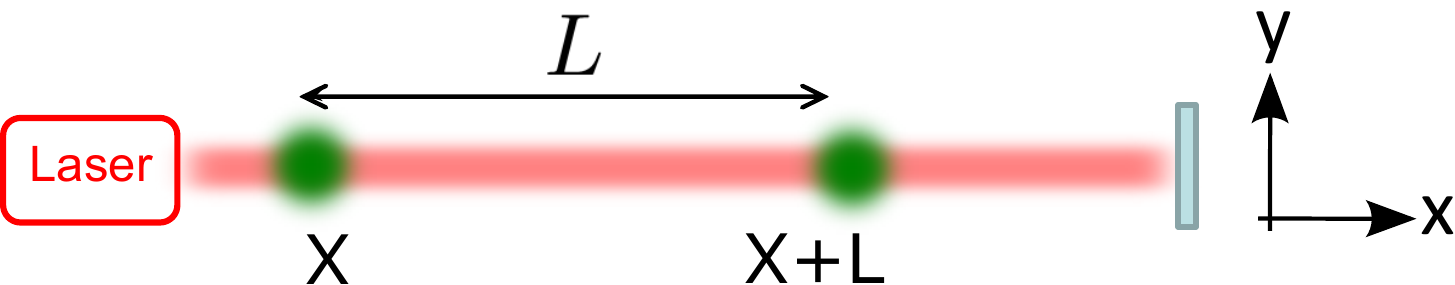}
\caption{Gradiometer configuration: two distant AIs (green clouds) are interrogated by the same laser beam.}
\label{fig:gradiometer}
\end{figure}

We now consider the gradiometer configuration sketched in Fig.~\ref{fig:gradiometer}.
Taking the differential signal $\psi(X,mT_c)=s_\varphi\left(X,mT_c\right)-s_\varphi\left(X+L,mT_c\right)$ 
between two AIs separated by the distance $L$  yields:
\beq
 \psi(X,mT_c) = 2k\left[L\left(\frac{\ddot{h}(t)}{2}-\frac{\delta\ddot{\nu}(t)}{\nu_0}\right)+\delta X\left(X+L,t\right)-\delta X\left(X,t\right)\right]\otimes s(t-mT_c).
\eeq
This equation can be rewritten in terms of the local gravity acceleration as
\beq
\label{eq:gradient}
 \psi(X,mT_c) = 2k\left[L\left(\frac{\ddot{h}(t)}{2}-\frac{\delta\ddot{\nu}(t)}{\nu_0}\right)+a_x\left(X+L,t\right)-a_x\left(X,t\right)\right]\otimes s_\alpha(t-mT_c)
\eeq
where $a_x(X,t)=\partial_t^2 [\delta x(X,t)]$ is the local gravity acceleration in the $x$ direction and $s_\alpha(t)$ is the AI sensitivity function to acceleration, given by $\ddot{s}_\alpha(t)=s(t)$.

The very important aspect in this equation is that the position noise $\delta x_2(t)$ of the retroreflecting mirror has been rejected by the gradiometer configuration. To be more precise, position fluctuations of frequencies smaller that $(2L/c)^{-1}$ are rejected, which represents the major part of the position noise in optical GW detectors (see Chapter \ref{sec:ch4}).
Rejection of the vibration noise in gradiometer configuration has already been measured in AIs (rejection by 140 dB was demonstrated in \cite{McGuirk2002}). This important immunity to position noise of the AI gradiometer  makes such instruments good candidates for GW detectors operating at lower frequencies than ground based optical interferometers, which sensitivity are limited at frequencies below $\sim 10 \Hz$ by position noise of the optics (vibration noise, thermal noise, etc) (see Chapter \ref{sec:ch4}).

Eq.~\eqref{eq:gradient} also shows that  fluctuations of the local gravity field result in an acceleration signal $a_x(X,t)$ whose gradient will have the same signature as that of the GW. Therefore, it is impossible to distinguish the effect of a GW from that of a fluctuating gravity gradient. This fundamental limitation is known as the gravity gradient noise limit, or Newtonian Noise limit, and has been the subject of several studies in ground-based optical GW detectors \cite{Saulson1984}. We will explore the possibility to reduce the Newtonian Noise with AIs in section \ref{sec:rejection_NN}.

\subsection{Quantum limited strain sensitivity curve}
\label{sec:strain_sensitivty}
To illustrate the potential performance of the AI detector, we will assume in this paragraph that the detector is limited by the quantum noise, i.e.  we neglect in particular the contribution of laser frequency noise and Newtonian noise which appear in Eq.~\eqref{eq:gradient}.
The power spectral density (PSD) of the gradiometer output is then given by
\beq
S_\psi(\omega) = (2nk L)^2   \omega^4 \frac{S_h(\omega)}{4} |\hat{s}_\alpha(\omega)|^2  +  2 S_\phi(\omega),
\label{eq:sensitivity_curve}
\eeq
where $S_h$ is the PSD of the GW, $S_\phi$ is the PSD of the AI phase noise (the factor 2 accounts for the 2 AIs involved in the gradiometer), and $\hat{s}_\alpha(\omega)=\text{FT}[s_\alpha(t)]=4\sin^2\left(\omega T/2\right)/\omega^2$ is the Fourier transform of the AI sensitivity function to acceleration.

The factor $n$ in Eq.~\eqref{eq:sensitivity_curve} denotes the number of momenta transfered to the atom during the diffraction process, which amplifies the phase signal by a factor of $n$. It is analogous to the use of Fabry-Perot cavities which amplify the phase signal in laser interferometers. Such process called large momentum transfer (LMT) beam splitters  is now frequently used in AI experiments to enhance the sensitivity of the interferometer (see, e.g. \cite{Clade2009,Chiow2011}). Proof of principle AIs with $n=100$ have been reported.

The phase noise PSD for an AI limited by quantum noise can be written as
\beq
S_\phi(\omega)=\frac{\eta}{\dot{N}_{at}} \ \left(\frac{\text{rad}^2}{\text{Hz}}\right)
\eeq
where $\dot{N}_{at}$ is the cold atom flux (in s$^{-1}$) and $\eta\leq 1$ is a factor which accounts for a possible measurement noise reduction with respect to the standard quantum limit ($\eta\approx 0.01$ has ben reported in \cite{Hosten2016}).

\begin{figure}[!h]
\centering
\includegraphics[width=\linewidth]{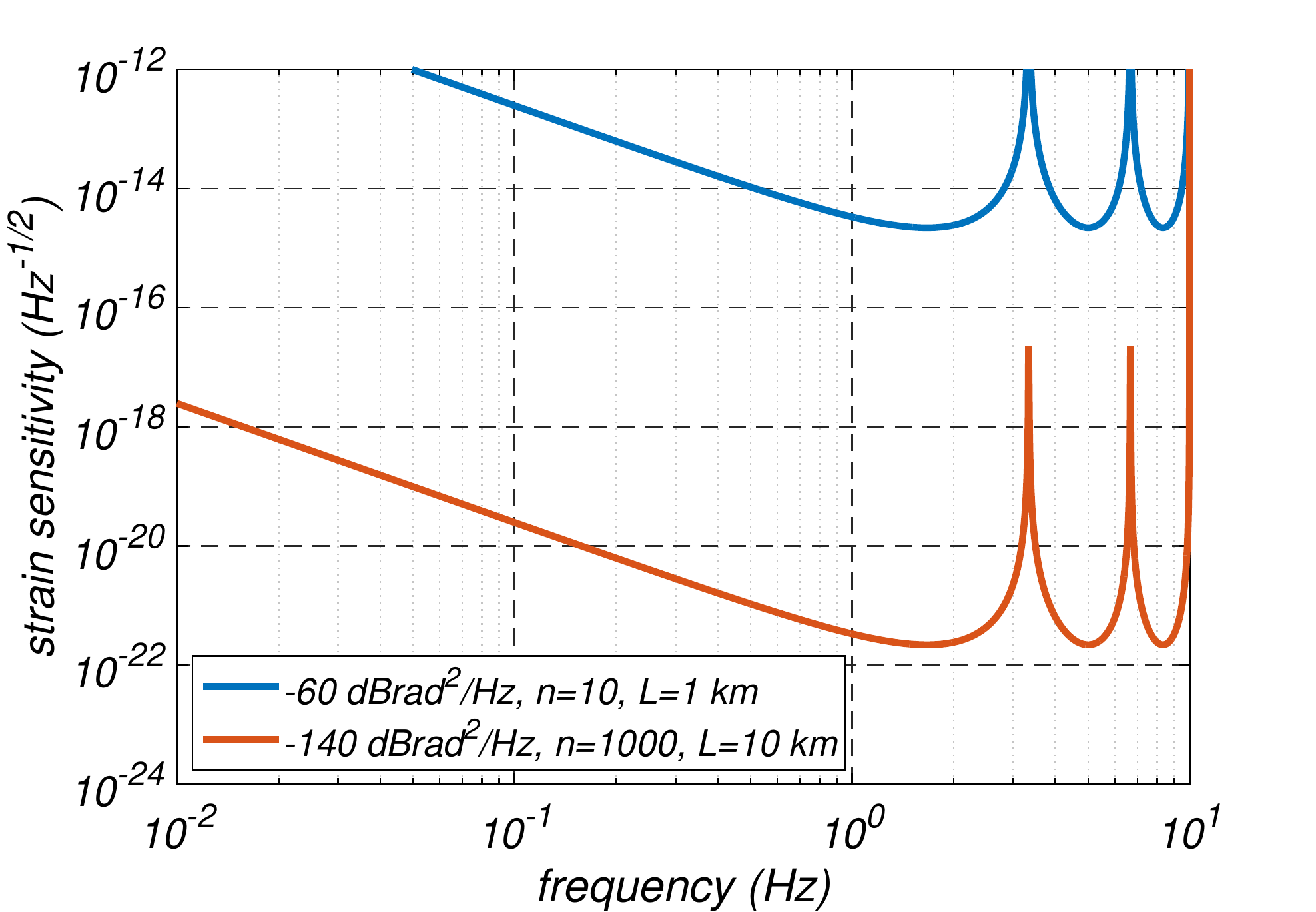}
\caption{Quantum limited strain sensitivity curve of the AI gradiometer GW detector for different parameters of the AI. In both cases $T=0.3$~s.}
\label{fig:sensitivity_function}
\end{figure}

If we consider a minimum sensitivity with a signal to noise ratio of 1, we obtain the strain sensitivity function:
\beq
\left(S_h\left(\omega\right)\right)^{1/2} = \left(\frac{2\eta}{\dot{N}_{at}}\right)^{1/2}\frac{1}{4Lnk\sin^2\left(\omega T/2\right)}.
\label{eq:sensitivity_curve}
\eeq
We plot in Fig.~\ref{fig:sensitivity_function} the strain sensitivity function for various parameters of the AI gradiometer. 
The blue line  corresponds to an optimized  AI combining several state of the art techniques, i.e. with a phase noise of $10^{-6}\  \frac{\text{rad}^2}{\text{Hz}}$ and a 20 photon LMT  beam splitter ($n=10$), and a gradiometer baseline $L=1$~km.  The red curve corresponds to the much more ambitious scenario which could be obtained in the future with $10^{-14}\  \frac{\text{rad}^2}{\text{Hz}}$ phase noise, $n=1000$ and $L=10$~km. 
In both cases, we considered an interrogation time $2T=0.6$~s, which determines the frequency $f_0=1/2T$ corresponding to  the best sensitivity. Such interrogation time is typical and would allow to cover the frequency band $\sim 0.1-10 \Hz$. 
Using longer interrogation times $T$ does not change the value of the peak sensitivity but shifts the operating bandwith to lower frequencies. Long ($>10 \ \text{s}$) interrogation times in AIs using ultracold atoms (temperature $<10 \ \text{nK}$) could then be used to design space-based detectors operating in the mHz regime\cite{Dimopoulos2008PRD}.
Finally, we neglected here the sequential operation of the AI, i.e. we neglected the possible aliasing effects due to the finite sampling period $T_c$ of the AI.  We will discuss  this approximation in the context of increasing the AI bandwith at the end of the chapter, section \ref{sec:challenges}.

\section{Experimental techniques}
\label{sec:experimental_techniques}
In this section, we briefly present the main experimental techniques for realizing a cold atom interferometer. Various type of AIs exist; we present here the architecture which has led to the most significant results in the field, and which is currently mostly considered for applications to GW detection.

\paragraph{Why cold atoms ?}
Cold atoms are necessary to realize the light-pulse AI sensitive to inertial effects presented in section \ref{sec:light_pulse_AI}. The requirement on the atom temperature comes from the interrogation time of the atoms in the interferometer, in the range of hundreds of milliseconds. For an atom cloud with rms velocity $\sigma_v$, its radius after a time $2T$ of free propagation  is $\sigma_r\simeq 2 \sigma_v T$, which must be kept smaller than the interrogation laser beam radius. For a 1 cm waist laser beam, this condition places a bound on the rms velocity in the cm/s regime, corresponding to $\mu K$ temperatures (depending on the mass of the atom). Besides such transverse selection (atoms escaping transversaly from the laser beam), the frequency of the laser is Doppler shifted depending on the atom velocities. If the Doppler width $k\eff\sigma_v$ of the  distribution is greater than the Rabi frequency  $\Omega/2\pi$ of the two-photon transition, then some atoms will not undergo the transition, which will reduce the contrast of the AI. For typical laser powers corresponding to $\Omega/2\pi\sim 50 \kHz$, the velocity selection condition also requires atom temperatures in the $\mu K$ regime.

\begin{figure}[!h]
\centering
\includegraphics[width=\linewidth]{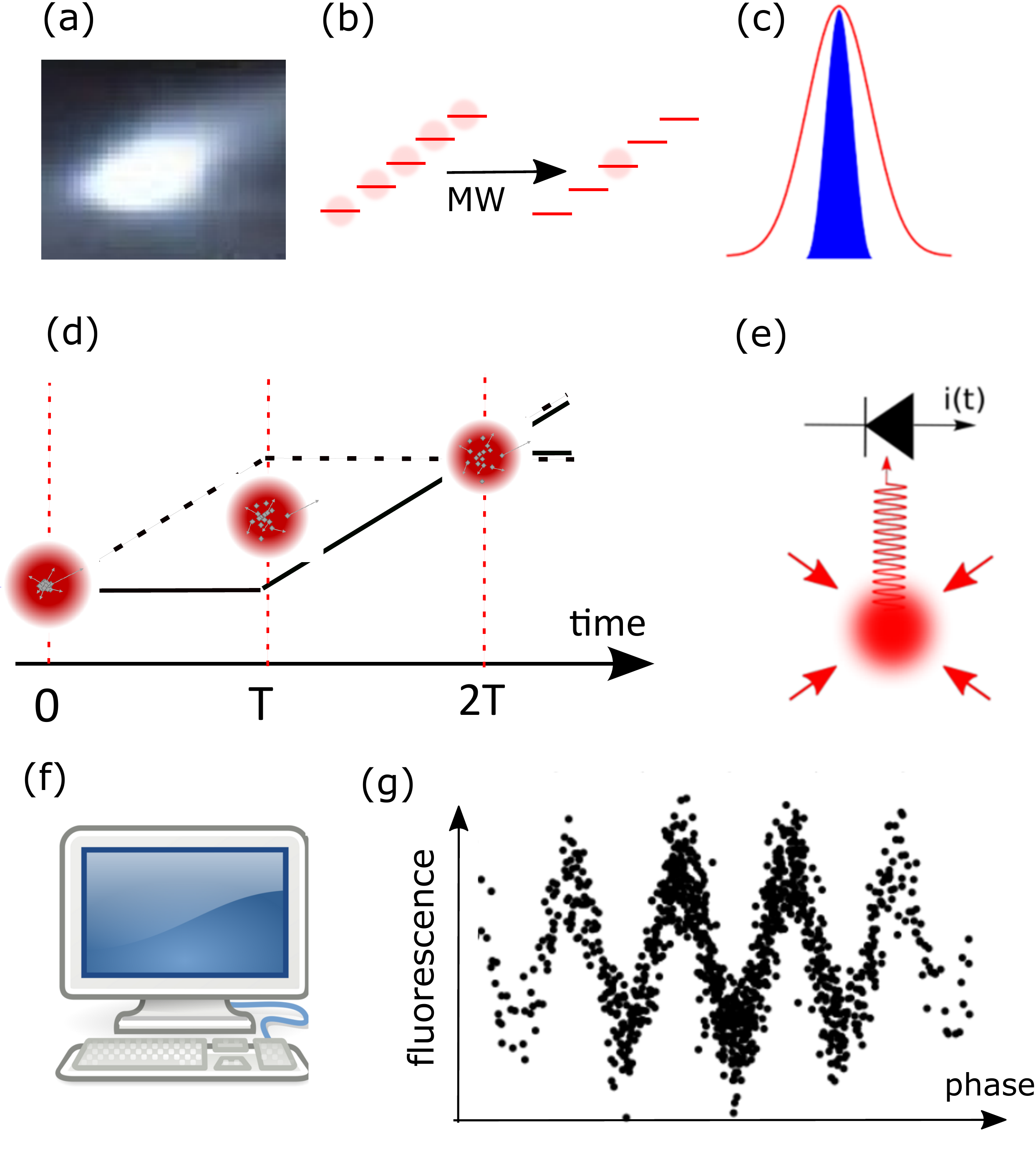}
\caption{Typical AI experimental sequence. (a) Laser cooling of the atoms in a Magneto-Optical-Trap. (b) Selection of the $m_F=0$ Zeman sublevel. (c) Longitudinal velocity selection (only the blue distribution will participate to the AI). (d) Light-pulse AI sequence. (e) Fluorescence detection of the AI output atoms. (f) Data analysis. (g) Interference fringes where each point correspond to one experimental cycle.}
\label{fig:exp_sequence}
\end{figure}

\paragraph{Experimental sequence.}
The typical experimental sequence is sketched in Fig.~\ref{fig:exp_sequence}.
It starts by laser cooling the atoms in an ultra-high vacuum chamber (pressure $\sim 10^{-10}$~mBar) (a). A detailed and pedagogical description of the associated physics and experimental techniques is presented in \cite{Metcalf2001}. Alkaline atoms have first been preferred because of their relatively simple electronic structure and of available laser sources.  For example, Rubidium 87  has extensively been used in cold atom experiments, because of the availability  of laser sources at 780 nm to cool this atom and manipulate its quantum states with electric dipole transitions. The laser cooling  step last typically hundreds of milliseconds to one second and produces $10^8-10^9$ atoms at a temperature close to $1 \ \mu K$ in a volume of few mm$^3$. 
The atom source is then further prepared before enterring the AI region, in order to improve its coherence (b) and (c): this includes, for example, a selection of the atoms in the $m_F=0$ Zeeman sublevel which is less sensitive to magnetic fields, and a velocity selection light pulse to enter the interferometer with a narrow (few 100 nK) longitudinal velocity distribution (in the direction of the AI lasers). As a result, about $10^6-10^7$ useful atoms participate to the interferometer (d). 
At the AI output, the population of atoms in each port is measured, for example by fluorscence detection with photodiodes (e). This allows to reconstruct the probablity of transition of the atom from the AI input state to one of the output states (f). This transition probability is modulated by the AI phase (g).

\section{Main noise sources and mitigation strategies}
As was shown at the end of section \ref{sec:phase_shift}, position noise of the optics is mainly rejected in the AI gradiometer configuration, which represents an important advantage of atom interferometry compared to laser interferometry. However, several other noise sources will affect a detector based on AIs. We review in this section some of the main noise sources. 
We focus on two important noise sources which are common to laser and atom interferometry: laser frequency noise and Newtonian noise. We present the strategies which were proposed to reduce these noise sources in detectors based on AI. 

\subsection{Strategies for the rejection of laser frequency noise}
\label{sec:laser_noise}
As shown in Eq.~\ref{eq:gradient}, the effect of the GW is indistinguishable from a fluctuation of the interrogation laser frequency at the same frequency: this is encoded in the term $L(h/2+\delta\nu/\nu)$. The origin of the sensitivity to laser frequency noise comes from the propagation delay between the two counterpropagating lasers originating from different locations. In the retroreflecing configuration which we considered (Fig. ~\ref{fig:scheme_lasers}), the extra phase accumulated by the beam which travels to the mirror and reflects back to the atoms is $2k(L-X_1)$ (see Eq. ~\eqref{eq:phi_P}), yielding a fluctuation $2\delta k (L-X_1)$ if the laser frequency fluctuates by $\delta k =2\pi\delta \nu/c$. For the other AI situated at a different location $X_2$ with respect to the retroreflecting mirror, the effect of the laser noise is different. Therefore, in the gradiometer configuration, the effect of laser noise amounts $2\delta k (X_2-X_1)$. The influence of laser frequency noise in such light-pulse AIs has been measureed in \cite{LeGouet2007EPJD}.  
Aiming at strain sensitivities of $10^{-20}\sqrtHz$ or lower in the frequency band $\sim 0.1-10 \Hz$ requires a laser with a relative frequency stability better than this level, or dedicated strategies for laser noise rejection/reduction.


\paragraph{Multiple arm configuration.}
A first possibility to reduce the effect of laser frequency noise is to adopt a cross arm configuration \cite{Dimopoulos2008PRD,Harms2013,Chaibi2016}, as in laser Michelson interferometric GW detectors.
Considering a symmetric configuration consisting of 2 orthogonal arms of same length and interrogated by the same laser, laser frequency noise is  rejected for a GW with (+) polarization. 
In laser interferometers, the degree of assymetry between the 2 arms sets the rejection efficiency for laser noise, which is typically  $99\%$(see Chapter \ref{sec:ch4})and limited by the assymetry bewteen the optical modes resonating in the two cavities. To achieve the required stability, the laser is therefore stabilized on the common mode of the Michelson interferometer, i.e. on the ultrastable km-long cavities inside the interferometer arm. This allows to reach a relative frequency stability $\sim 10^{-21}\sqrtHz$ in the detector frequency band, thus a strain sensitivity $\sim 0.01\times 10^{-21}\sqrtHz$.
The rejection of laser frequency noise in a cross arm detector using AIs has been discussed briefly in  Ref.~\cite{Chaibi2016}, but the rejection efficiency in various geometries has not been analyzed in enough details yet.

\paragraph{Single laser AIs.}
In 2011, Yu and Tinto  propose an AI detector configuration based on a single laser operating two distant AIs, instead of the two lasers considered so far   \cite{YuTinto2011}. Their proposal exploits coherent superpositions of two atomic levels separated by an energy corresponding to an optical transition, as in optical clocks. In such a scheme, the momentum transfer is performed by a single transition between two levels,  instead of a two-photon transition involving 3 levels as in the case of Fig.~\ref{fig:diffraction} which we considered so far.
Atoms characterized by a long ($\sim$~second) lifetime of the optically exited state can be used for such a protocol, as it is the case for example for alkaline-Earth-like atoms  used in optical clocks (Calcium, Strontium, Ytterbium). As only one laser is used to drive the transition, the problem explained above of laser noise sensitivity due to the propagration delay in a \textit{single} AI disapears. Another interpretation is to say that the AI compares the phase of the laser against the atomic internal clock coherence \cite{YuTinto2011}. 

Yu and Tinto conclude their article by considerations on the sensitivty of their proposed detector, and highlight the need to use large momentum transfer techniques (LMT, obtained by multiphoton transitions) to enhance the sensitivity of the detector. Their  idea is extended for the LMT configuration and  further detailed in \cite{Graham2013} in the context of a  space-based detector.
The authors claim favorable sensitivities at milli-Hertz frequencies compared to space-based laser interferometers such as LISA. 
The complexity of such a mission is  discussed in \cite{Bender2014}, in particular the need for very large laser powers required to drive the highly forbidden optical transition between the two (optical) clock state. To conclude this paragraph, we note that a Strontium AI using LMT beam splitter has been reported in 2015 \cite{Mazzoni2015}, and that many developments of AI with ultracold alkaline-Earth-like atoms are currently carried out by several research groups. We can expect a promising future for this technology.

\subsection{Rejection of the gravity gradient noise with an array of atom interferometers}
\label{sec:rejection_NN}
As shown by Eq.~\ref{eq:gradient}, the effect of the GW cannot be distinguished from that of a fluctuating gravity gradient by using two AIs. This problem is similar to optical GW detectors which use two test masses (the two cavity mirrors of one  interferometer arm) to probe the effect of the GW. This fundamental limit for GW detectors operating on Earth in known as the Newtonian Noise (NN) limit. For ground based detectors, it represents a fundamental limit which prevents from observing GW at frequencies below few Hz, because the NN starts to dominate at these frequencies \cite{Saulson1984,Beccaria1998a}. 

NN originates from mass fluctuations in the surrounding of the detector, which translate in gravity field fluctuations at the test masses. Sources of NN are, for example, seismic noise triggerring  stochastic fluctuations of the ground density and resulting in fluctuations of the gravity field (so called seismic NN), or air density fluctuations in the atmosphere caused by turbulence (so called infrasound NN).

\begin{figure}[!h]
\centering
\includegraphics[width=\linewidth]{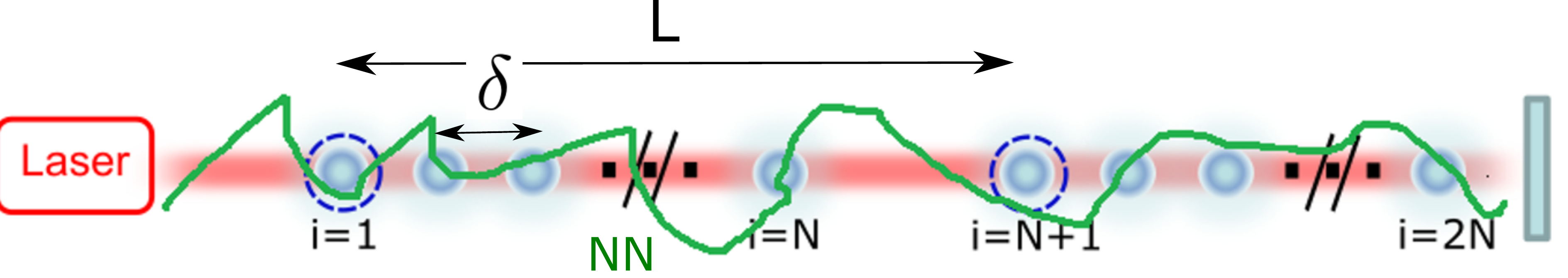}
\caption{Principle of the rejection of the Newtonian Noise (NN) with an array of spatially distributed AIs. $L$ is the gradiometer baseline and $\delta$ is the spatial sampling period.}
\label{fig:NN_rejection}
\end{figure}

An article published in 2016 presents a method to go beyond the NN limit in GW detectors based on atom interferometry \cite{Chaibi2016}. The method relies on the fact that the spatial properties of the NN are different  than the spatial properties of the GW: while the wavelength $c/f$ of the GW at $f=1 \Hz$ is $3\times 10^8$~m, the characteristic length $v/2f$ of the NN at such frequency is of order 1 km \cite{Saulson1984} ($v$ is the velocity of seismic waves for the seismic NN, or of the sound in air for infrasound NN). Therefore, by operating an array of spatially distributed AIs interrogated by the same laser beam, it is possible to average the NN to zero. 
 This idea is sketched in Fig.~\ref{fig:NN_rejection}.

More precisely, the detector consists of $N$ gradiometers of baseline $L$, and which sample the NN with a spatial step $\delta$. The average signal
\beq
H_N\left(t\right) =\frac{1}{N}\sum^{N}_{i=1} \psi(t,X_i),
\label{eq:average}
\eeq
 with the $\psi(X)$ signal given by Eq.~\eqref{eq:gradient}. This procedure yields the GW signal and a residue of the NN which standard deviation is reduced by $\sqrt{N}$ compared to the single gradiometer case, if the $N$ measurements are uncorrelated.
Using the spatial behavior of the NN correlation function, the authors show that a rejection of the NN greater than $\sqrt{N}$ can be obtained. 
For a scenario with $N=80$, it is shown that rejection efficiencies of up to 30 can be achieved  at 1 Hz.

\subsection{Comparison with other GW detectors}
\label{sec:comparison_other_det}

\begin{figure}[!h]
\centering
\includegraphics[width=\linewidth]{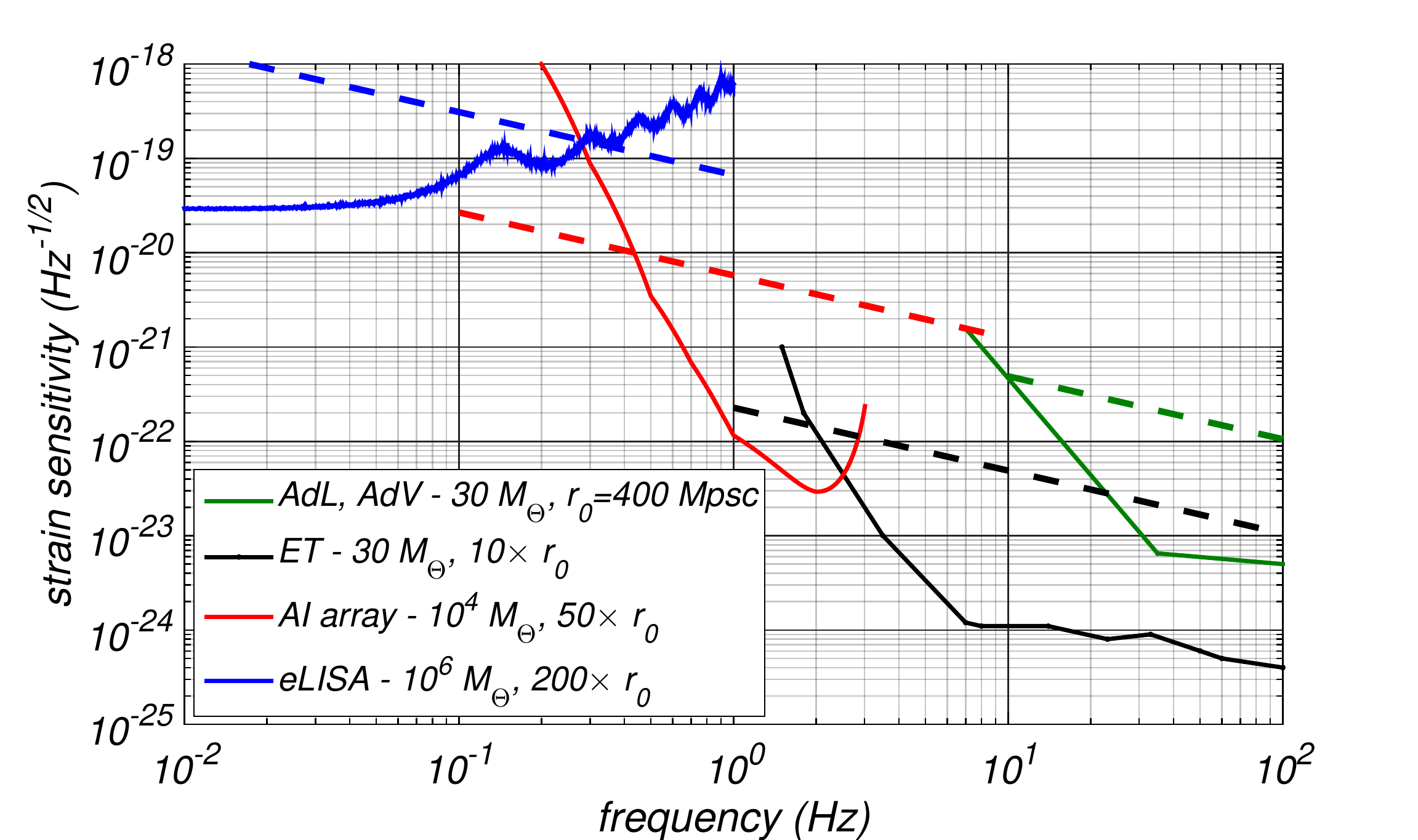}
\caption{Strain sensitivity curves of current  and possible future GW detectors (plain lines). 
Dashed lines indicate the required senitivity to detect the GW from a compact binary following Eq.~\eqref{eq:required_sensitivity}; $M_\Theta$ is the mass of the sun and $r_0=400 \ Mpsc$. We assumed a detection with a signal to noise ratio of 1. 
AdV: Advanced VIRGO \cite{AdV2014}, AdL: Advanced LIGO \cite{AdL2015}; ET (Einstein Telescope)  is a ground-based laser interferometer proposal \cite{Punturo2010}. eLISA \cite{eLisa2012} is a space-based laser interferometer proposal. The AI array is a ground-based AI proposal \cite{Chaibi2016}.}
\label{fig:comparison_other_det}
\end{figure}

The window opened by the AI array proposition described in the previous section  would enable to cover a frequency band where no other detector is currenlty planned to operate, as shown in Fig.~\ref{fig:comparison_other_det}. 
The parameters of the AI detector (red line) are those of Ref.~\cite{Chaibi2016}: a phase noise level of $10^{-7} \ \text{rad}\sqrtHz$, LMT beam splitters with $n=1000$ and a detector baseline $L=16 \ \text{km}$. The quantum noise is moreover reduced by a factor $\sqrt{N}$ thanks to the array of $N$ AI.
In  Fig.~\ref{fig:comparison_other_det}, we  show   strain sensitivity functions for different detectors (plain lines) and simple estimates for the GW signal (dashed lines) corresponding to compact binaries as sources of the GW. We parametrized the binaries by the mass of the stars and their luminosity distance, and assumed a detection with signal to noise ratio of 1. To obtain the estimate of the signal strength for the compact binary, we followed the simple model of \cite{Harms2013} [Eq.(11) to (18)] yielding
\medskip
\beq
\sqrt{\left[S_h(\omega)\right]}\simeq 0.3 \times \eta^{1/2} c^{-3/2} f^{-2/3} (GM)^{5/6} r^{-1}
\label{eq:required_sensitivity}
\eeq
with $\eta=m_1 m_2 /M^2$ the symmetric mass ratio ($M=m_1+m_2$), $f$ the GW frequency and $r$ the luminosity distance.
We refer to Ref.~\cite{Sathyaprakash2009} and chapter  \ref{sec:ch2} for more details on the sources.

\subsection{Other noise sources}
\label{sec:other_backgrounds}
Several noise sources already identified in atom interferometry experiments will potentially affect the sensitivity of the GW detector. While the gradiometer configuration gives immunity to some of the noise sources because they are common to both AIs, several other backgrounds might affect the sensitivity and will depend on the exact nature of the detector, e.g. two-photon transitions or single photon transitions  to realize the diffraction. Some of them have been analyzed in \cite{Dimopoulos2008PRD,MIGA_NJP_in_prep} for the 2-photon transition case which we focused on in this chapter, and in \cite{Graham2013,Bender2014} for the single photon case. To cite few, we can mention the effect of wavefront distorsions of the laser beam which might be seen differently by the two distant AIs, magnetic field fluctuations, effects of rotations on the interferometer, residual effect of vibrations or laser phase noise.

\section{Current projects and perpsectives}

While several research groups worldwide are developing new atom interferometry techniques, and  studying theoretically the application to GW detection, only few teams started to realize an instrument which could address this application.
High sensitivity AIs using LMT techniques, ultracold atoms and long interrogation times in tall vaccum chambers are being constructed in USA \cite{Dickerson2013}, China \cite{Zhou2011}, Australia \cite{ANUwebsite} and Europe \cite{Hartwig2015}. 
However, hybrid laser-atom interferometers based on the   gradiometer configuration, which is primarily considered for application to GW detection, require long baseline instruments.
We present here the only project which, to date and to our knowledge, is being pursued towards the application to GW detetction with a long ($>100$~m) instrument: the Matter-wave laser Interferometric Gravitation Antenna (MIGA) project currently under construction in Europe.

\subsection{The Matter-wave laser Interferometric Gravitation Antenna (MIGA) project}
\label{sec:MIGA}
The MIGA project started in 2013 with an initial funding from the French nation research agency (ANR), and involves about 15 institutes with expertise in atomic physics,  metrology, gravitational physics and geosciences. The goal of the project is to design and realize an instrument capable of serving as a demonstrator for a future GW antenna based on atom interferometry. The instrument will also be used for precision gravity field measurements, with important applications in geosciences, in particular hydrology \cite{Geiger2015}.
The initial idea of the instrument is based on the gradiometer configuration described in section \ref{sec:gradiometer}, with a baseline $L=300$~m, and the possibility to correlate several AIs interrogated by the same laser beam.
Details on the design and  realization of the subsystems of MIGA  can be found in \cite{Canuel2014,Geiger2015,MIGA_NJP_in_prep}. We briefly describe here the main elements of the instrument, which commisionning should start in 2018.

\begin{figure}[!h]
\centering
\includegraphics[width=\linewidth]{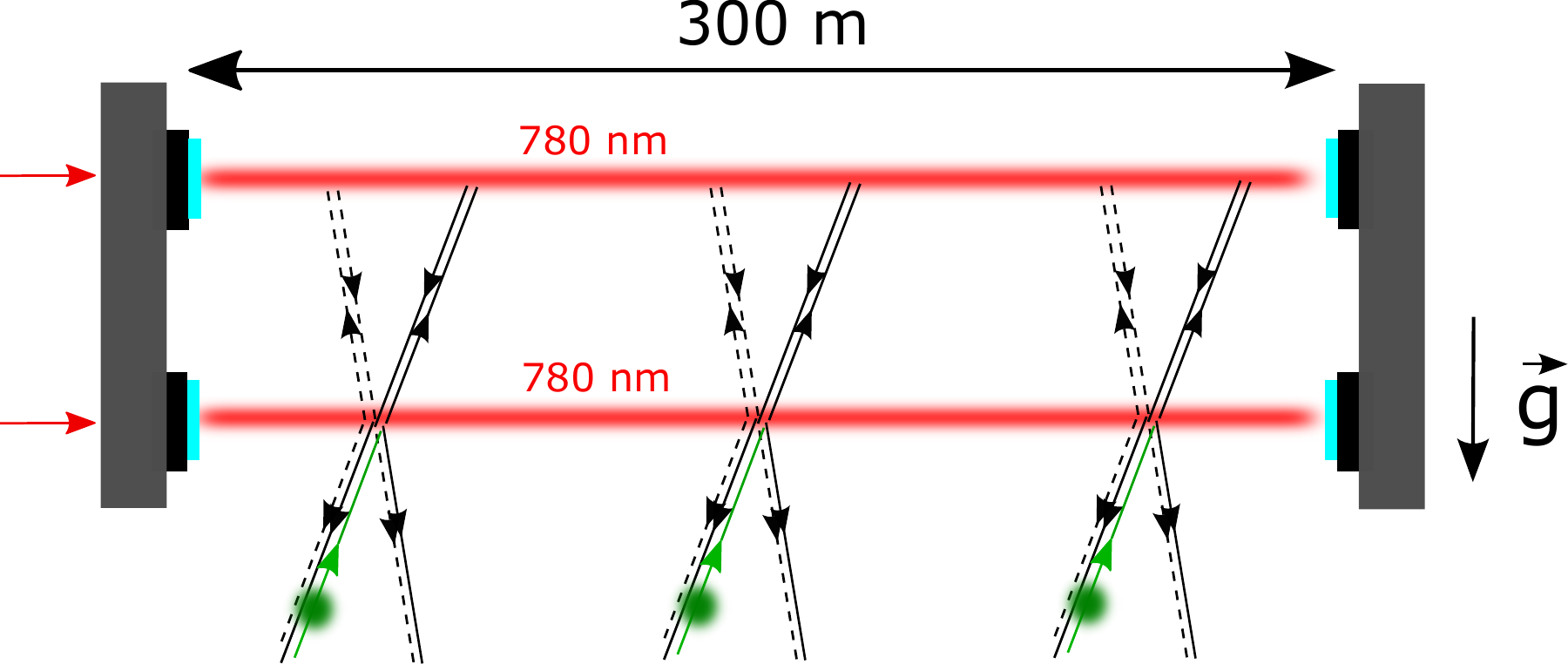}
\caption{Geometry of the MIGA instrument (not to scale, the separation between the two AI arms being of order of few cm). The atoms are launched from below from the magneto-optical trap (not shown). Bragg diffraction on lasers resonating in an optical cavity is used as matter-wave optics, which requires to enter the AI with the correct angle. The Bragg angle ($\sim$~mrad) is exagerataed here for clarity. The interrogation time is $2T=500$~ms for a 3 light-pulse AI with apogee of the atom trajectory at the top beam.}
\label{fig:MIGA_geometry}
\end{figure}

Fig.~\ref{fig:MIGA_geometry} shows the geometry of the MIGA instrument where the optical mode of two optical cavities  interrogate simultaneously 3 AIs seperated by a distance of about 150 m.  The optical cavity will allow to enhance the optical power at the cavity resonance by the optical gain ($\simeq 10$ in the initial design) in order to improve the efficiency of LMT beam splitters which require large laser powers. The length of the cavity will be stabilized using a 1560 nm wavelength laser stabilized at the $10^{-15}$ relative frequency stability. 
The AIs use Bragg diffraction of $^{87}$Rb on the light standing wave in the cavity (wavelength 780 nm), where momentum states $|-n\hbar k\rangle$ and $|+n\hbar k\rangle$ are coupled by the (high order) Bragg diffraction. LMT beam splitters with $n\simeq 5$ are initially planned for atom sources with temperature in the $\mu K$ range.
In this figure, the cold atom source is not shown. Atoms are cooled in a magneto-optical trap (MOT) located about 1 meter below the first Bragg beam, and launched vertically at a velocity close to 5 m/s. On its  way up, the atom source is prepared as mentionned in section \ref{sec:experimental_techniques}. After the AI, on the way down, the atomic state is probed by fluorescence detection, which allows to measure the interferometer phase shift.

\begin{figure}[!h]
\centering
\includegraphics[width=\linewidth]{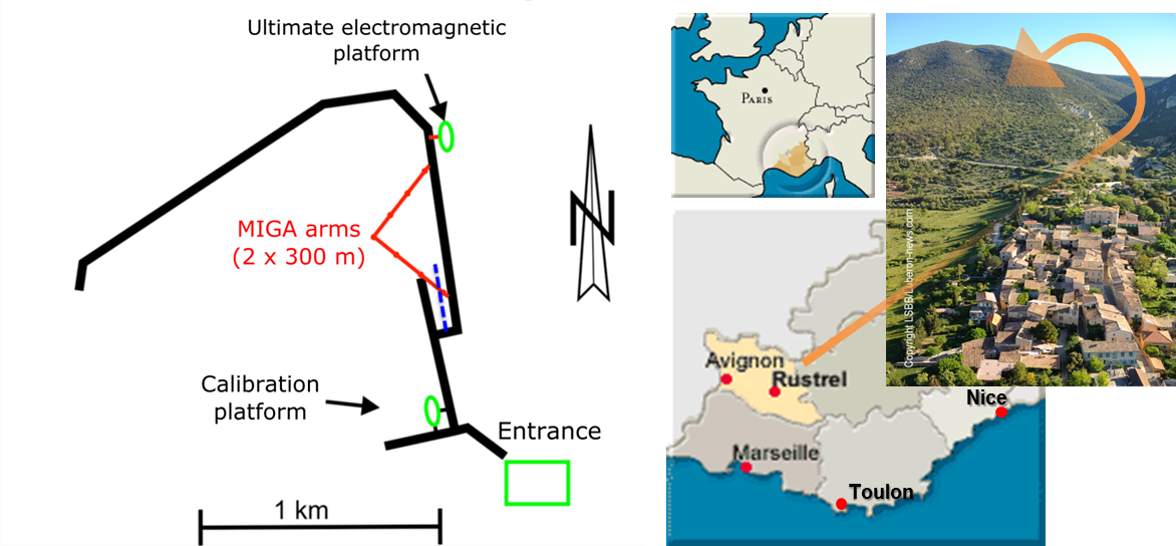}
\caption{Installation site of the MIGA instrument at the low noise underground laboratory in the South-East of France.}
\label{fig:rustrel}
\end{figure}

The instrument will be installed at the low noise underground laboratory LSBB loacted in the South-East of France \cite{LSBBwebsite}, see Fig.~\ref{fig:rustrel}. Two 300 meter galleries will be dug dedicatedly for the detector. Besides the vacumm tube, the optical systems and the AI sensors, various environmental instruments will be deployed in order to monitor the environment around the detector and assess applications in hydrology: superconducting gravimeters,  seismometers, radars or muon detectors.

While the MIGA instrument has not been designed for GW detection applications because of its relatively small (300 m) baseline and the lack of maturity of cold atom technology at the start of the project, it will serve as a first demonstrator for a future larger detector, at the European or international scale. Among the first tests which will be performed, correlations between distant AI sensors in an optical cavity, gravity gradient measurements at the $10^{-13}$~s$^{-2}\sqrtHz$ level, and validation of the state of the art AI technology in a large detector will have an important impact on the design of a future instrument.

\subsection{Challenges}
\label{sec:challenges}
As shown in section \ref{sec:strain_sensitivty}, reaching strain sensitivities of order $10^{-22}\sqrtHz$  in the $\sim 0.1-10 \Hz$ frequency band will require many technology developments. The technology should progress by about 10 orders of magnitude in strain sensitivity to be fully useful for GW astronomy. Many of the required techniques have been demonstrated independently in various experiments. 
Given the rapid progress in the field in the last 20 years, we can thus expect breakthroughs in the next decades. 
On the basis of the detector design presented in this chapter, future instruments will probably require 
\begin{enumerate}
\item LMT beam splitting techniques \cite{Mueller2008,Clade2009,Chiow2011,McDonald2013};
\item ultracold atoms  \cite{Muttinga2013,Kovachy2015a};
\item quantum enhanced phase measurements  \cite{Hosten2016} and/or much brigther cold atom sources \cite{Robins2013,Bolpasi2014,Raizen2014}
\item  higher sampling rates/continuous operation of the AI \cite{Meunier2014,Dutta2016}. 
\end{enumerate}
The challenge will consist in realizing an instrument which combines all these techniques. The table below summarizes the requirements, as well as the state of the art.
A new concept of detector based on different ideas than that presented here would allow to revisit these numbers and require less ambitious values.

\begin{table}[h!]
\centering
\tbl{Technology requirements for a GW detector based on a AI gradiometer configuration to reach a strain sensitivity of $10^{-22}\sqrtHz$. The "current" column corresponds to demonstrated results in different experiments. The "required" column implies that all  techniques are operationnal in the same experiment.}
{\begin{tabular}{c||c|c|c|}
\toprule
AI parameter & current & required  \\ \hline \hline  \\
LMT order ($n$) & 100 & 1000  \\
temperature$^{\text a}$ & 1 nK & ? \\
interrogation time $T$ (s) & $0.1-1$ & $\sim 0.3$ \\
AI operating frequency (Hz) & 5 & 20 \\
phase sensitivity (dBrad$\sqrtHz$) & -30 & -70   \\
detector baseline  & 300 m & $>3$ km \\ \hline
strain sensitivity ($\sqrtHz$) & $10^{-10}$ & $10^{-22}$  \\ \hline \hline 
\end{tabular}}
\begin{tabnote}
$^{\text a}$Low temperatures are required for high contrast AIs based on LMT diffraction, but this requirement depends on the available power of the laser which drives the LMT pulse.
\end{tabnote}
\label{tab:parameters}
\end{table}

We conclude this section by comparing the signal of the AI gradiometer $\Psi\stxt{gradio}$ to that associated with the effect of the GW on a single AI, $\Delta\phi$, given by eq.~\eqref{eq:phase_shift2}. The ratio between the two contributions is 
\beq
\frac{\Psi\stxt{gradio}}{\Delta\phi}\sim \frac{L n k}{T v_L n k} =\frac{L}{v_L T}\approx \frac{10^4 \ \text{m}}{10 \ \text{m.s}^{-1} \times 1 \ \text{s}}=10^3,
\eeq
It shows that the contibution from the differential laser phase imprinted on the AI dominates over the GW induced phase on the atomic wave in a single AI. This estimation should however be revisited in details.

\section{Conclusion}
\label{sec:conclusion}
GW astronomy will benefit from the largest frequency band covered by different detectors. While the performance of current ground based laser interferometers is impressive and triggered GW astronomy, their sensitivity at frequencies below  10 Hz might strongly be limited. It is therefore important to look for complementary solutions.
One possible solution is to use cold atoms as test masses to probe the phase of the laser influenced by the GW. Such solutions started to be studied in the 2000' after the rapid progress of the field of atom interferometry.
This is the possibility which has been described in this chapter. 

GW detectors based on AI rely on probing the  phase of a laser with free falling cold atoms, and are characterized by an important immunity to  position noise of the optics. Moreover, strategies have been proposed to reduce some of the noise sources identified in laser interferometric detectors, such as laser frequency noise or gravity gradient noise. 
In this chapter, we presented strategies for designing a ground based GW detector sensitive in the $\sim 0.1-10 \Hz$ frequency band.
Besides the principle of a detector based on atom interferometry, we presented the current projects which have started to  design of a future GW detector, as well as the many technological challenges that remain.

The schemes which were initially studied and abandonned because of experimental complexity should be reconsidered, taking into account the important technological progress in the field of atomic physics.
In particular the possibility to use a single AI based on relativistic particles for GW detection should be revisited.

\section{Acknowledgements}
I would like to thank  my collaborators from the atom interferometry and inertial sensor team of the 
SYRTE laboratory and from the MIGA consortium.
I thank Pac\^ome Delva and Christian Bord\'e for their useful comments on the historical aspects of the introduction.

\bibliographystyle{../ws-rv-van}    
\bibliography{../Geiger/cold_atoms_biblio-1}      


%